# Identification of Free and Bound Exciton States and Their Phase-Dependent Trapping Behavior in Lead Halide Perovskites

Jiangjian Shi[1], Huiyin Zhang[1, 2], Yiming Li[1, 2], Jacek J. Jasieniak[3*], Yusheng Li[1, 2], Huijue Wu[1], Yanhong Luo[1, 2], Dongmei Li[1, 2], Qingbo Meng[1, 2*]

*1. Key Laboratory for Renewable Energy, Chinese Academy of Sciences; Beijing Key Laboratory for New Energy Materials and Devices; Institute of Physics, Chinese Academy of Sciences, Beijing 100190, P. R. China*

*2. University of Chinese Academy of Sciences, Beijing 100049, P. R. China*

*3. Department of Materials Science and Engineering, Monash University, Clayton, Victoria 3800, Australia*

** Corresponding author: qbmeng@iphy.ac.cn, jacek.jasieniak@monash.edu*

**Abstract:** In this work we probe the sub-gap energy states within polycrystalline and single crystal lead halide perovskites to better understand their intrinsic photophysics behaviors. Through combined temperature and intensity-dependent optical measurements, we reveal the existence of both free and bound exciton contributions within the sub-gap energy state manifold. The trapping and recombination dynamics of these excitons is shown to be strongly dependent on the structural phase of the perovskite. The orthorhombic phase exhibits ultrafast exciton trapping and distinct trap emission, while the tetragonal phase gives low monomolecular recombination velocity and capture cross-sections ($\sim 10^{-18}$ cm$^2$). Within the multiphonon transition scenario, this suppression in charge trapping is caused by the increase in the charge capture activation energy due to the reduction in electron-lattice interactions, which can be the origin for the unexpected long carrier lifetime in these material systems.

Hybrid and inorganic metal halide perovskites have rapidly emerged as promising materials for use within high-efficiency photovoltaic,[1-5] light-emitting,[6-7] and detecting[8] devices. These successes have arisen because of the outstanding optoelectronic properties that such materials possess, including high light absorption coefficients of $>10^5$ $cm^{-1}$ across the visible region,[9-10] long carrier lifetimes that can exceed 1 μs,[11-13] and unexpectedly slow hot-carrier cooling processes.[14-15] Underlying these alluring characteristics is an intrinsic photo-physical mechanism that dictates the absorption of light, thermalization of carriers, carrier inversion, and recombination or charge-transfer dynamics.[16-17] For perovskite materials, these intrinsic phenomena are yet to be fully understood.

Recent studies have highlighted the importance of a hot-phonon bottleneck and polaron screening effects towards explaining the excited state carrier dynamics in these materials.[14-15] Following thermalization, reports have suggested that free carrier formation dominates at higher temperatures,[18-19] while excitonic contributions, which include resonance,[20] localization,[21] and transfer,[22] also contribute to the photophysics. Consistent with traditional inorganic semiconductors, these exciton states have been attributed to the band-edge light absorption of perovskites;[23] however, contrasting reports have also suggested that these arise from band-tail states[24] or indirect band transitions.[25] Notably, an indirect bandgap model has most recently been invoked to describe the band-edge bimolecular recombination in these systems.[26] Despite these controversies, it is evident that low-energy electronic states within the direct bandgap (i.e. sub-gap states) of perovskites play a fundamental role in defining their optoelectronic properties.

Sub-gap states in semiconductors can act as intermediate energy levels to facilitate charge recombination or as bound states suitable for charge localization.[27] In perovskites, the importance of such states is confirmed with studies showing that trap-mediated monomolecular processes dominate the recombination dynamics,[19,25] while also accelerating ion migration and performance degradation.[28-30] The importance of these factors ensues that a comprehensive understanding into the role of sub-gap states, especially those of excitonic or carrier-trap origins, and their related photophysics, is critical towards further improving perovskite materials and their device performances.

Herein, we study the sub-gap state related absorption and fluorescence characteristics of methyl ammonium lead bromide ($MAPbBr_3$) perovskites to better understand their photo-physical origins. Through temperature and time-dependent measurements, we find explicit signatures of a free and bound exciton in both polycrystalline and single crystal samples. The spectral dynamics of these excitons exhibit phase dependent behavior, with the orthorhombic-to-tetragonal phase transition effectively prohibiting exciton trapping. Within the multiphonon transition theory, this behavior originates from an increase of the trap capture activation energy following the phase transformation. These findings provide a clearer understanding into the origins of the photo-physical properties and related charge carrier processes within perovskite materials.

## Results

### Free and bound excitons in the perovskite polycrystalline.

Studies on the low-energy absorption of perovskites have shown that a bathochromic shift occurs when the perovskite is grown from a polycrystalline thin-film to a single crystal.[23, 31] The diversity of exciton states at the crystal surface and in the bulk have been suggested to be the origin for this phenomenon.[23] However, these propositions have not adequately considered the potential variation in defect density, lattice ordering, and perovskite thickness across these samples, which can readily obscure the underlying cause of any absorption edge changes. To exclude these effects, we have selectively used polycrystalline samples of a common origin for this study. $MAPbBr_3$ was chosen as the representative perovskite sample because its absorption edge and emission are in the visible-region, and its orthorhombic-to-tetragonal phase transition occurs at a sufficiently low temperature [21] of ~140 K to decouple phase-related and intrinsic characteristics. The $MAPbBr_3$ polycrystalline were grown by direct precipitation within a viscous ethylacetate (EA)/ethylcellulose (EC) solution to form a concentrated colloidal perovskite dispersion. This dispersion could be readily diluted through additional EA/EC; thus, enabling the perovskite concentration to be continuously adjusted, while preserving its underlying physical and chemical nature.

Measurements of the absorption and PL characteristics of these perovskite dispersions at different concentrations are shown in Figure 1(a) and 1(b), respectively. Both exhibit progressively more prominent low-energy optical features at the higher concentrations. An analysis of this low-energy absorption edge as a function of perovskite concentration reveals a linear relationship (Figure 1c), which is consistent with the Beer-Lambert Law.[33] The PL characteristics can be readily explained by considering photon re-absorption effects arising from the modified absorption properties of the samples. Re-absorption has been previously considered to explain the PL spectrum of perovskite single crystals.[32] The impact of re-absorption on the PL can be estimated by considering that the PL Intensity ($I_{PL}$) detected from a sample is proportional to the product of the PL at a high dilution ($I_{PL0}$) and the light transmittance ($T$) between the excitation focal point and detector: $I_{PL}(E) \approx I_{PL0}(E) \times T(E)$. The good agreement between the measured and estimated PL peak energies obtained at each concentration across our samples (Figure 1c), confirms that re-absorption is the primary cause of the PL spectral shift observed in our perovskite samples and not from the emergence of new emission states or the change of perovskite properties with concentration. This conclusion is further supported by PL decay measurements, which show very similar decay characteristics regardless of the perovskite concentration (Figure 1(d)). These simply state that the extended absorption edge in these samples is arising from electronic states that are intrinsic to each perovskite crystal in the sample.

Exciton, band-tail, and indirect-band states can all act as sub-gap energy states that contribute to the optoelectronic properties at the absorption edge. [16] To quantify these contributions, we first considered the conventional Tauc formalism,[26] which is composed of indirect ($E_{ig}$) and direct ($E_g$) bandgap terms scaled by proportionality factors $a$ and $b$, respectively: $\alpha(\mathrm{E})=a\times(E-E_{ig})^2+b\times(E-E_g)^{1/2}$. However, this approach proved unrealistic for perovskites, predicting them to be indirect semiconductors despite it being well known that that they are direct in nature (Supplementary Figure 1). A more suitable approach to extract the absorption-related energy states involved the Elliott formula, which considers the continuum-band and excitonic contributions concurrently:[34]

$$\alpha(E) = \left[A \cdot \theta(E - E_g) \cdot D_{CV}(E)\right] \cdot \left[\frac{\pi x e^{\pi x}}{\sinh(\pi x)}\right] + A \cdot R_x \cdot \sum_{n=1}^{\infty} \frac{4\pi}{n^3} \cdot \delta\left(E - E_g + R_x / n^2\right) \qquad (1)$$

where $\alpha$ represents the light absorption coefficient, $A$ is a constant, $E$ is the photon energy, $\theta$ is the step function, $D_{CV}$ is the joint density of states described as $D_{CV}\sim(E-E_g)^{1/2}$ near the direct band edge, $n$ is the principal quantum number of the exciton state, $\delta$ represents a delta function, and $x = [R_x/(E-E_g)]^{1/2}$, with $R_x$ being the exciton binding energy.

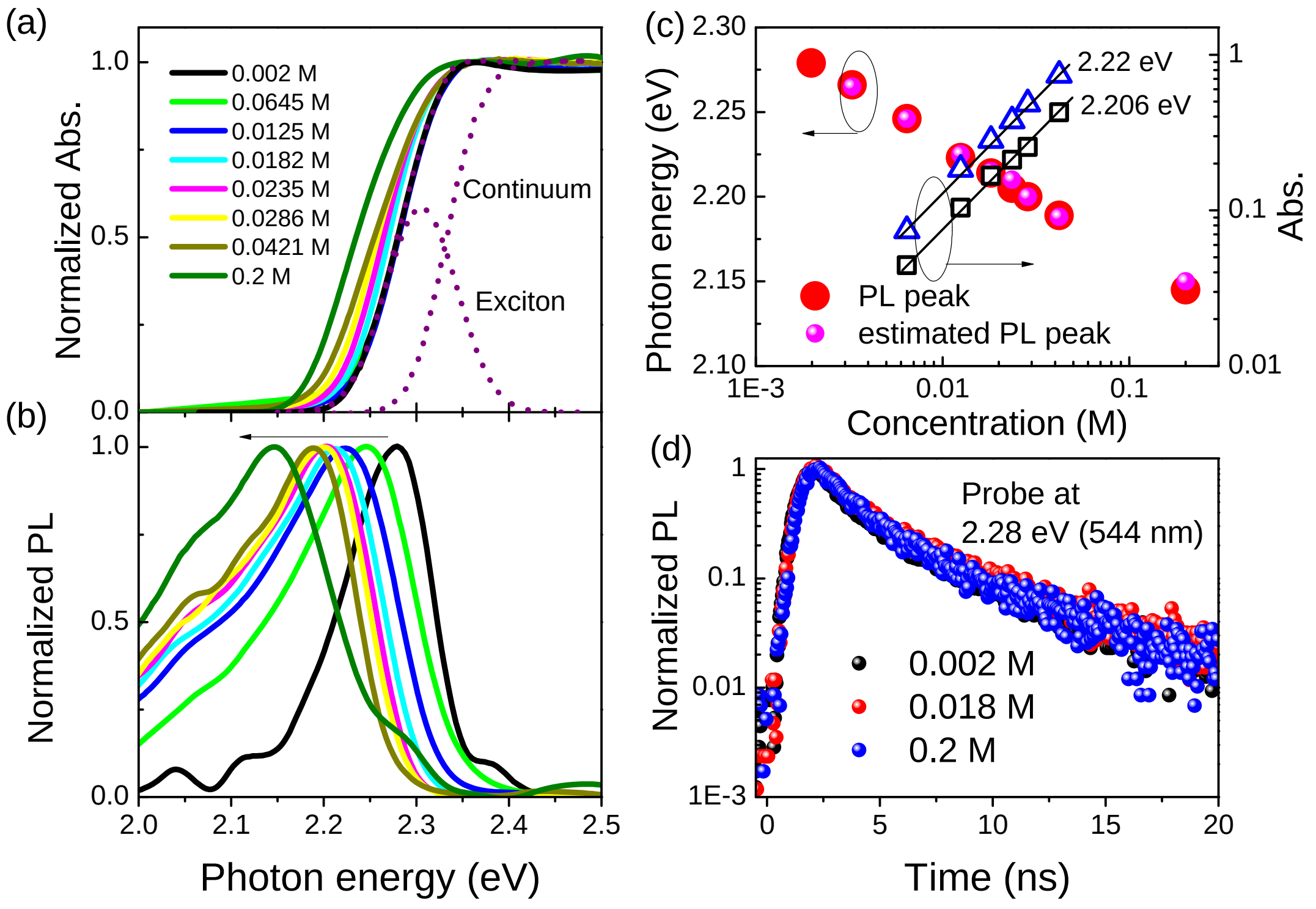

**Figure 1. Light absorption and emission properties of the $MAPbBr_3$ colloidal solutions at different concentrations.** (a) Normalized light absorption (the absorption spectra of the low-concentrated perovskite is fitted with a continuum-band and exciton model) and (b) steady-state PL spectra of the perovskite slurries. (c) The measured and estimated PL peak position (left axis) and the low-energy light absorption (right axis) as a function of the concentration. (d) Transient PL of the perovskites probed at 2.28 eV.

Using the Elliot formalism, with a Gaussian broaden exciton contribution, the absorption spectrum of the 2 mM perovskite is fitted to yielded an $E_g$ of 2.32 eV and an $R_x$ of ~20 meV (Figure 2(a)). This $E_g$ is similar to that reported by Wu et al. for single $MAPbBr_3$ crystals, while the $R_x$ is much smaller than the ~60 meV found.[23] Considering the thermal energy at room temperature is ~26 meV, an exciton with an binding energy of only 20 meV can be considered to exist as a dissociated or free exciton. For the 0.2 M perovskite, evidently the absorption is extended to lower energy, but it cannot be adequately fitted by just considering the continuum-band and single exciton transition (Supplementary Figure 2). Instead, we find that the Elliot formalism modified to include a secondary, low-energy Gaussian-shaped absorption band that is centered at 2.245 eV provides an excellent fit to the experimental results (Figure 2(b)). Assuming a constant continuum and exciton contribution between the dilute and concentrated samples, we determine an exciton binding energy for this additional resonance of ~ 75 meV.

The absorption tail below 2.22 eV can also be described by the indirect bandgap model (Supplementary Figure 2), which yields an $E_{ig}$ of 2.138 eV. However, this model cannot provide a satisfactory global fit to the absorption spectra and is not consistent with the PL results as presented in the following sections. In addition, because of the ultra-low absorption coefficient in the tail region,[24-25] the exponential band tail (Urbach model) also cannot be used to explain the absorption characteristics here. Based on the above observations, we argue that the emerging low-energy absorption contribution in our $MAPbBr_3$ originates from another exciton state, which has a higher binding energy of up to 75 meV and, thus, can be considered a bound exciton at room temperature.

In Figure 2(c)-(g) we show fitted absorption curves of all the samples with the continuum-band term subtracted for clarity. These results clearly show that the low-energy bound exciton absorption resonance becomes more evident at higher perovskite concentrations. Importantly, for all the perovskites, the free exciton resonances are located at the same position of ~ 2.30 eV and with a similar Gaussian broadening of ~ 38 meV. This is also the case for the bound exciton state, which always locates at ~2.25 eV with a broadening of ~ 33 meV. A similar behavior is observed for $MAPbI_3$ samples, with two exciton states of binding energy 17 meV and 67 meV being observed within the bandgap ($E_g$=1.645 eV, Supplementary Figure 3). Notably, investigations into single crystal perovskites have also implied multiple exciton contributions for which the low-energy (high energy) contribution was assigned to the bulk (surface) of the perovskite single crystal.[23] Our results suggest that these two types of exciton states are intrinsic to the perovskite material, no matter the surface or the bulk.

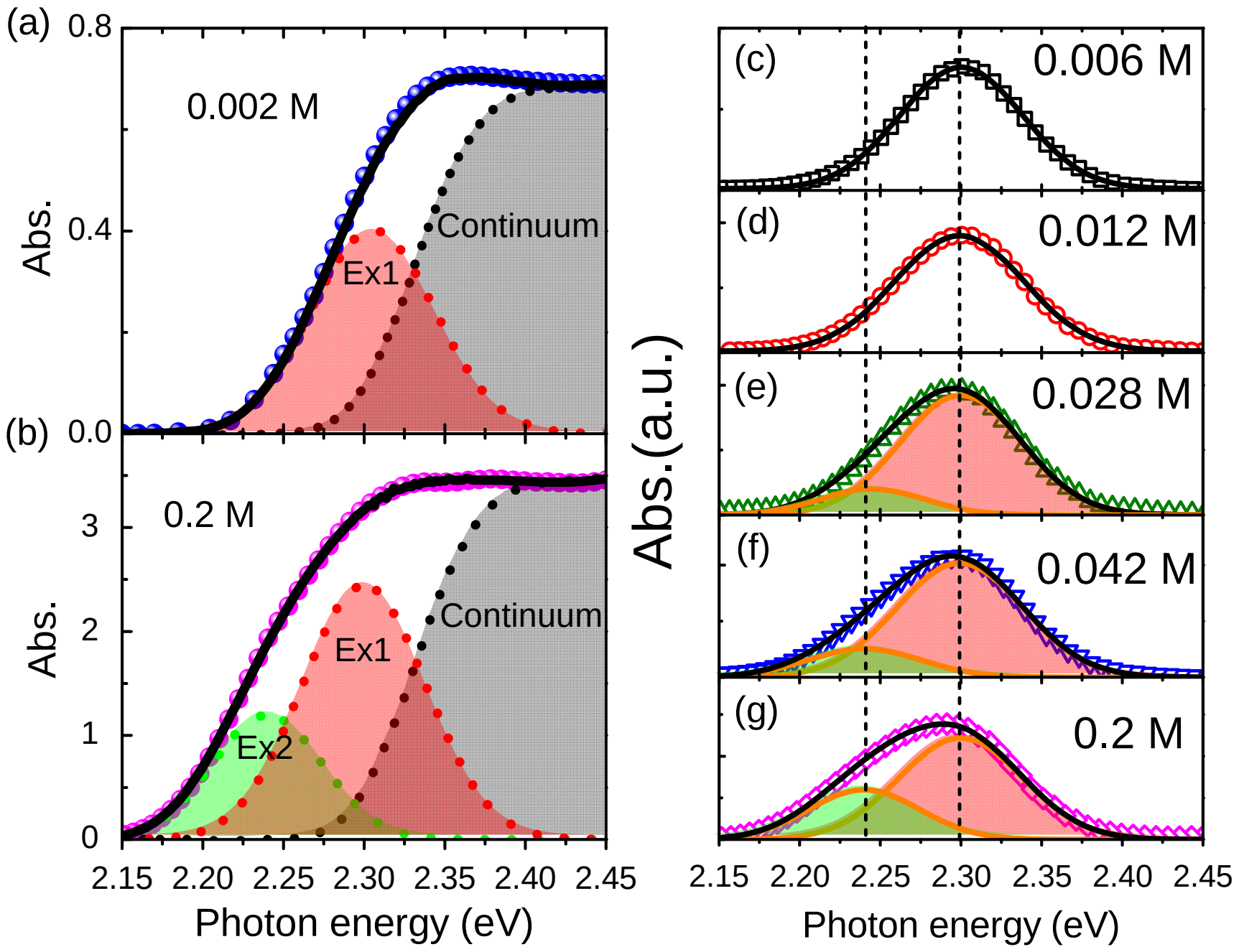

**Figure 2. Fits of perovskite light absorption spectra with a continuum-band and double exciton model.** (a) and (b) give the full fits for the colloidal samples with 0.002 M and 0.2 M perovskite concentration, respectively. (c-g) The evolution of the exciton absorption when increasing the perovskite concentration. For clarity, the continuum-band term has been subtracted from the spectra.

To further probe the underlying photo-physics of the $MAPbBr_3$ system, we have carried out steady-state PL on perovskite films as a function of temperature from 10 K to 300 K. At temperatures lower than 130 K, a sharp PL peak centered at about 550 nm is observed, along with a broad emission band centered at about 620 nm (Supplementary Figure 4). These can be assigned to exciton and trap emissions, respectively.[35-36] At higher temperatures, the trap emission disappears and the exciton emission is significantly enhanced. Moreover, at 145 K, two PL peaks located at 541 nm (2.292 eV) and 550 nm (2.255 eV) are distinguished in the PL spectrum (Figure 3(a)). The energies of these PL peaks are similar to that of the free and bound exciton states, respectively, which may imply the successful observation of free and bound exciton emission. Notably, the $MAPbI_3$, the PL contribution at room temperature is also composed of two bands, suggesting a similar origin (Supplementary Figure 3).

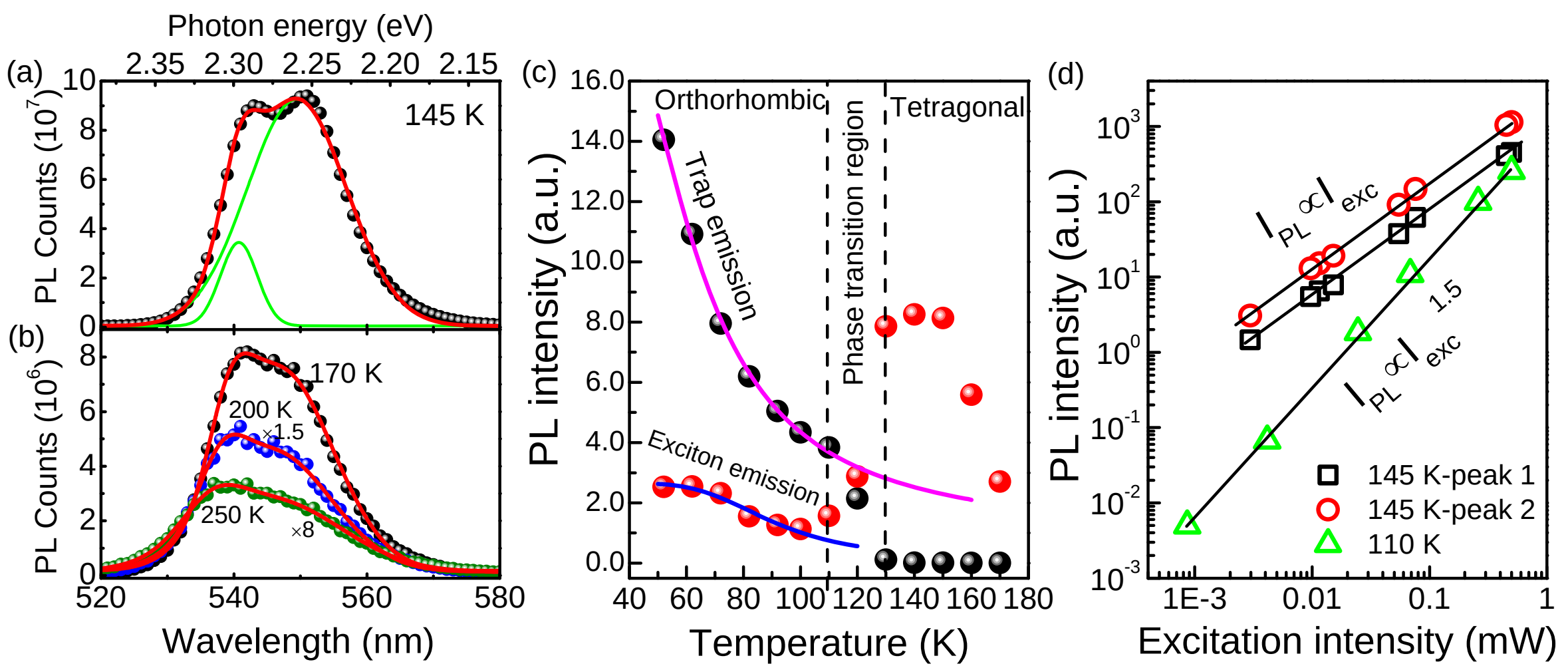

**Figure 3. Temperature-dependent steady-state PL spectra of the $MAPbBr_3$ films.** Experimentally measured and fitted PL spectra at (a) 145 K and (b) 170 K, 200 K and 250 K. (c) Temperature-dependent exciton (free and bound excitons represented as an average) and trap emission intensities. In the low temperature regime of < 120 K, the relationships are fitted with the thermal quenching model. (d) Excitation intensity ($I_{exc}$) dependent PL intensity ($I_{PL}$) of the perovskite at 110 K and 145 K, respectively. The data are fitted with $I_{PL} \sim I_{exc}^{c}$, which at 145 K gives $c \approx 1$ for each of the two emission peaks and at 110 K gives $c \approx 1.5$. For all the measurements, the perovskite was excited at 445 nm with an averaged excitation intensity of about 50 mW $cm^{-2}$.

In addition to surface and bulk origins,[23] it has been proposed that a double-peak emission in perovskites can result from a mixture of perovskite phases (orthorhombic and tetragonal)[35, 37] as well single and triplet state emission.[38] To probe the underlying origin in more detail, we measured the PL at much higher temperatures. The double-peak emission can be observed from 145 K all the way through to 250 K (Figure 3(b)); while for the $MAPbI_3$, this can even be observed at room temperature. At these high temperatures, phase pure samples should be expected. Further evidence for this is garnered through a comparison of the temperature ($T$)-dependent PL intensity ($I_{PL}(T)$) (Figure 3(c)). At a temperature lower than 110 K, both the intensities of the exciton and trap emissions decrease with temperature, following the Arrhenius thermal quenching model.[39] A dramatic decrease in the trap emission intensity is observed at temperatures beyond the 110 K, with a concurrent enhancement of the exciton emission being noted. This behavior deviates from the thermal quenching model, thus implying a change in the emission mechanism across the different temperature regimes, which is directly correlated to the internal phase of the material. Further analysis of the excitation intensity ($I_{exc}$) dependent PL measurements (Figure 3(d)) shows that at 110 K, the PL intensity changes according to the power relationship $I_{PL}$~$I_{exc}^{c}$ (ref. 40-41) with $c$≈1.5. However, at 145 K, a linear relationship (i.e. $c$ ≈1) is obtained for each of the two emission peaks, indicating a different charge transfer and dynamic process.[41] These trends in PL behavior suggest that the orthorhombic phase participating in the luminescence is completely transformed into the tetragonal phase at temperatures higher than 130 K. Notably, it has been previously reported that the multiple-peak PL observed at 50 K in such systems arises from triplet exciton emission;[38] however, based on the excitation intensity dependent PL peak blue-shift, it is more likely to arise from intrinsic trap states.

**Exciton at the surface and exposed bulk of perovskite single crystals.**

Polycrystalline perovskite powders and films possess inherent inhomogeneity due to the existence of grain boundaries and lattice disorder.[36] To better understand the origins of the multiple PL contributions, we have grown single crystal $MAPbBr_3$ by the inverse temperature crystallization method,[42] which preferentially exposes the cubic (100) plane (Supplementary Figure 5). We carefully cleaved the single crystal to expose the bulk lattice. Short-wavelength excitation within a reflection-mode collection geometry was used to measure the emission wavelength-dependent PL transients of both as-prepared and cleaved surface regions. The resulting two dimensional pseudocolor plots of these transient PL spectra are shown in Figure 4 for 165 K. Both samples show two emission centers located at about 543 nm (2.28 eV) and 554 nm (2.24 eV), which are consistent with that observed in the polycrystalline perovskite samples. This double-peak emission can also be observed when the temperature is increased to 200 K (Supplementary Figure 6). Notably, light-scattering contributions[32] were ruled out as a possible cause of this double-peak emission in the single crystal PL based on different decay rates for each emission band (Supplementary Figure 7). Thus, we can confirm that the free and bound exciton states also coexist in the perovskite single crystal.

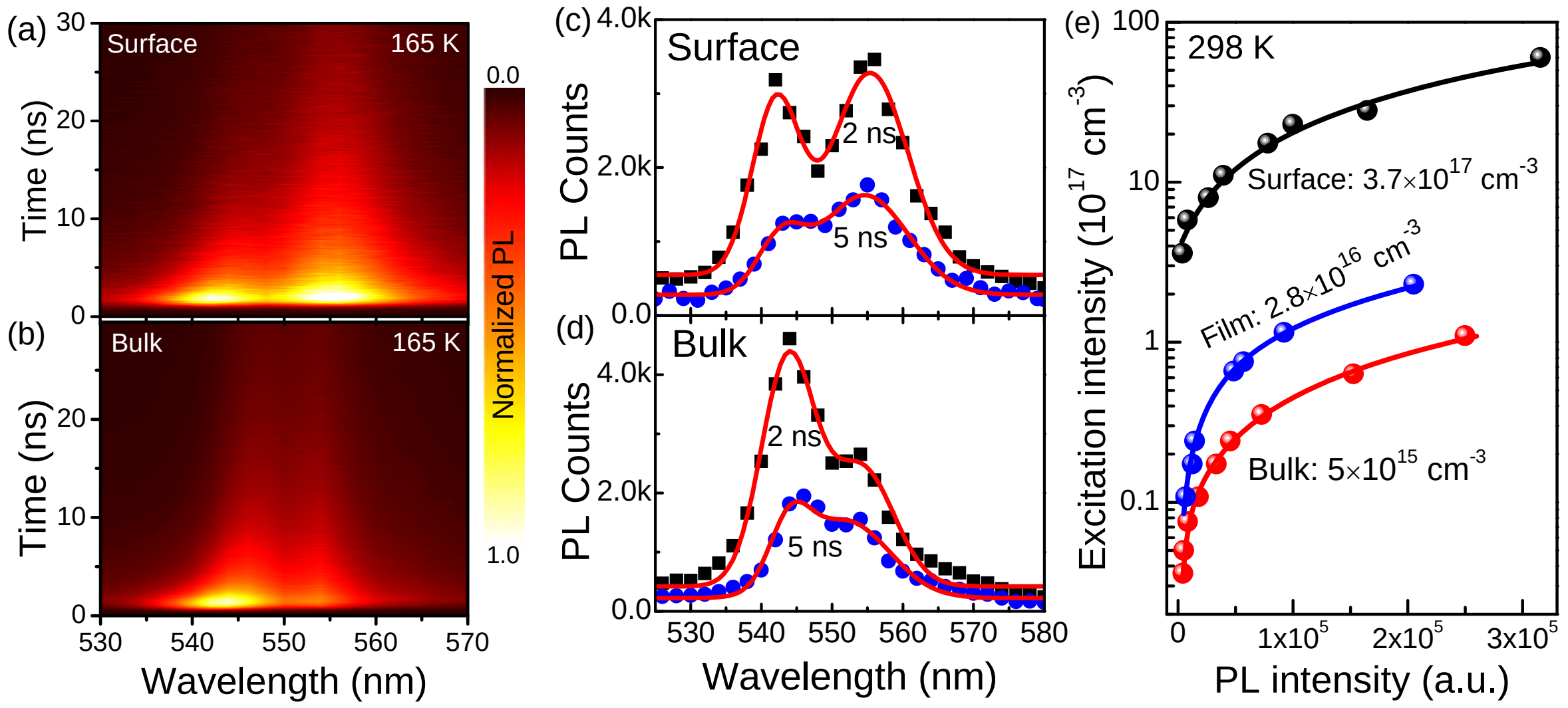

**Figure 4. PL of the perovskite single crystal.** Two dimensional pseudocolor (normalized PL intensity) plots of transient PL for the $MAPbBr_3$ single crystal as functions of emission wavelength and time delay (165 K, (a) surface and (b) bulk). The bulk is obtained by cleaving the single crystal and it is protected by a thin PMMA layer. Typical PL spectra of the perovskite single crystal at time delays of 2 and 5 ns, respectively ((c) surface and (d) bulk). These PL spectra are fitted with a double-peak emission model. (e) Trap density measurements of the perovskite film and single crystal (surface and bulk) by monitoring the pump density-dependent integrated PL intensity at room temperature.

The wavelength dependent PL spectra at different times following excitation have been extracted from the two dimensional pseudocolor plots (Figure 4 (c) and (d)). The free exciton emission contribution shows a slight red-shift in the early stage of the decay process, while the position of the bound exciton emission remains unchanged. This implies that the free exciton progressively transfers into lower energy states, likely of a localized nature,[21, 43] while the bound exciton is more energetically stable. Furthermore, it is observed that the relative PL intensity of the bound exciton for the single crystal surface is much higher than that for the bulk. Usually, a bound exciton is formed due to the coulomb interaction between an exciton and a trap center.[16] The greater existence of these states may imply that the trap density of the single crystal surface is much higher than that in the bulk.

To make such a correlation, the trap density of the perovskite single crystal (surface and bulk) and film is estimated using excitation intensity-dependent PL measurements. According to the semiconductor band-edge recombination model,[7,23] these experimental dependences can be fitted using $n(0)=N_t[1\text{-}\exp(-a\tau_0 I_{PL}/k)]+I_{PL}/k$, where $n(0)$ is the initial photocarrier density, $N_t$ is the trap density, $k$ is a constant for a certain sample, $\tau_0$ is the effective PL lifetime, and $a$ is the product of the trap capture cross section ($\sigma$) and the thermal velocity ($v_{th}$). To generate a relatively fast band-edge radiative rate for satisfying this model, this measurement was conducted at room temperature. Under this formalism, the trap density of the single crystal surface is estimated to be $3.7\times10^{17}$ $cm^{-3}$, which is about two orders of magnitude higher than that of the bulk ($5\times10^{15}$ $cm^{-3}$). This result agrees well with that reported by the two-photon measurement, [23] implying that the crystal quality (e.g. lattice ordering and defect) of the bulk perovskite have not been affected by the cleaving process. The correlation between the trap density and the bound exciton emission indicates that regulating the level of traps is necessary for charge localization for efficient perovskite emitting and lasing applications. Furthermore, a comparison of the PL characteristics derived from the perovskite films and single crystals, both at the surface and the exposed bulk, show similar behavior. This suggests that comparable energy states exist across the different perovskite samples we have studied.

Based on a PL lifetime for the polycrystalline perovskite films of ~ 350 ns (Supplementary Figure 9) at 298 K, $\sigma$ is estimated to be about $3.4\times10^{-18}$ $cm^2$. Comparable values of $\sigma$ were determined for other perovskite systems, such as the $MAPbI_3$ and (FA, MA)Pb(I, $Br)_3$. Impressively, these values are several orders of magnitude smaller than that for most traps in GaAs [44] and CdTe [45]. The small $\sigma$ explains the origin for the unexpected slow monomolecular recombination in this material, but the physics mechanisms behind this small value deserve further studies.

**Multiphonon transition-assisted exciton trapping and phase dependence.**

Despite the low $\sigma$, we find that the exciton can be rapidly captured by trap states at low temperatures. This can be readily observed at 70 K, for which three emission centers exist: free excitons, bound excitons, and trap mediated irradiative recombination (Figure 5(a)). For clarity, the time for PL maxima at different wavelengths are marked out by a black line and the transient PL probed at 545 nm and 620 nm is extracted and shown in Figure 5(b). The exciton emission states exhibits an instrument-limited fast rise and sharp drop behavior in the early stage; while for the trap emission, its rise process is much slower with no rapid drop being observed. A time delay of about 800 ps exists between these two types of emissions. The matching of the dynamics between the exciton emission decay and concomitant rise of the trap emission implies the possibility of ultrafast charge transfer from exciton to trap states, i.e. exciton trapping. Interestingly, at 170 K (tetragonal phase), the rapid drop of the exciton emission is no longer observed and the trap emission also disappears (Figure 5(b) and 3(c)). This behavior deviates from the thermal quenching model, where a higher temperature would accelerate the exciton transfer (Supplementary material note 1).[39] These results suggest that the phase transition prohibits the rapid trapping of the excitons at early time-scales. Notably, the hot carrier cooling process also exhibits a phase-dependent prohibition behavior,[15] implying a strong correlation between these two dynamics processes.

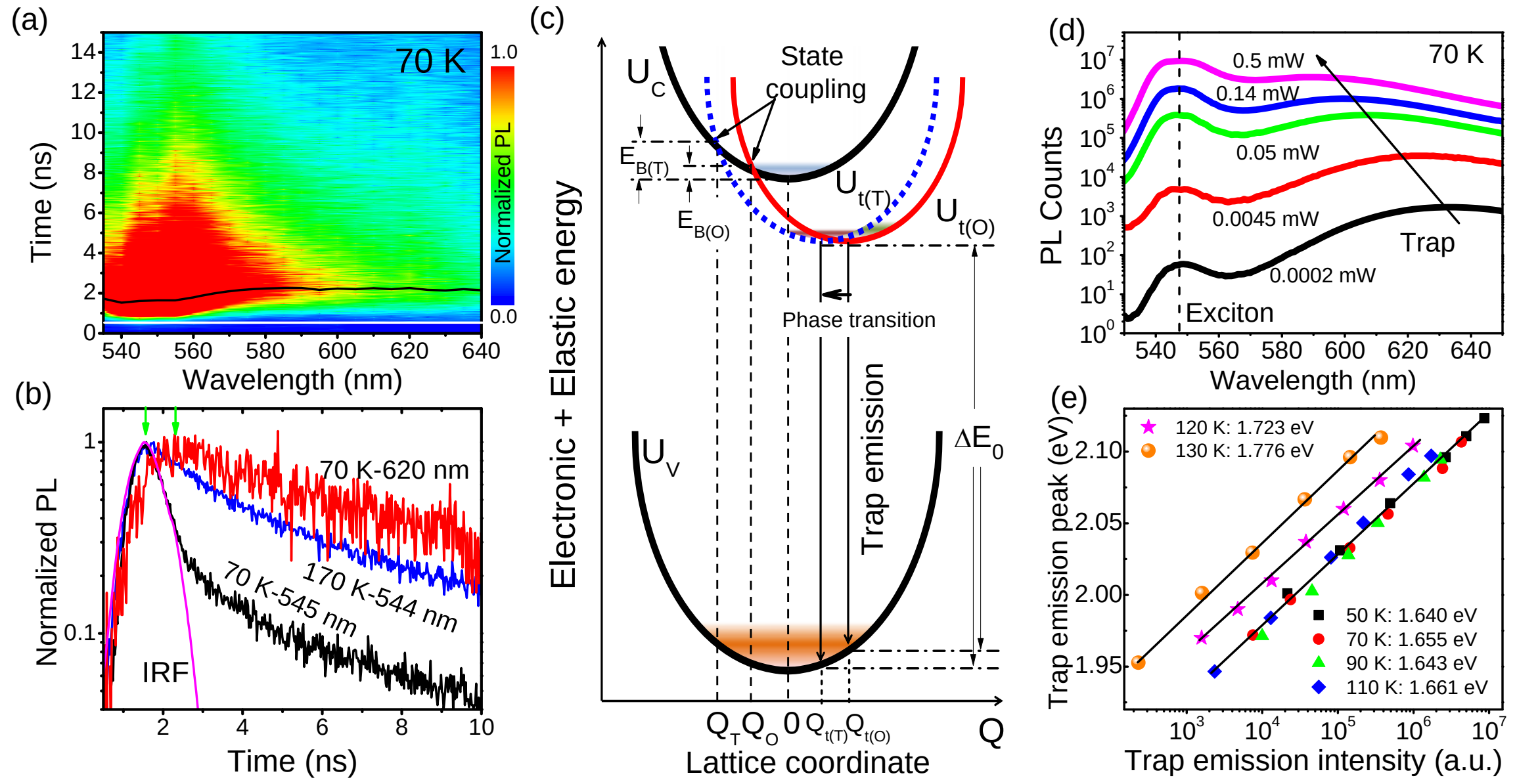

**Figure 5. Exciton capture and trap emission of the perovskite at low temperatures.** (a) Two dimensional pseudocolor (normalized PL intensity) plots of transient PL as functions of emission wavelength and time delay at 70 K. The black line marks out the time of the PL maximum at different wavelengths. (b) Transient PL of exciton and trap emission in the early stage. The green arrows show the time delay of exciton and trap emission. (c) Configuration coordinate diagram of total energy (electronic + elastic energies) of conduction band ($U_C$), trap state ($U_t$), and valence band ($U_V$) versus single lattice coordinate $Q$ in a semiconductor.

The transition from orthorhombic to tetragonal phase makes the trap state shift in the $Q$ space and moves the state crossing point from $Q_O$ to $Q_T$, thus increasing the charge capture activation energy from $E_{B(O)}$ to $E_{B(T)}$. The equilibrium position of the trap state is consequently moved from $Q_{t(O)}$ to $Q_{t(T)}$, which directly influences the energy difference for trap emission. (d) Excitation intensity-dependent steady-state PL spectra of the perovskite at 70 K. Two peaks corresponding to the exciton and trap emission are marked. The trap emission shows a progressive blue-shift under higher excitation intensities. (e) Logarithmic relationship between the trap PL peak and the intensity at temperatures ranging from 50 to 130 K. The minimum energy difference for the trap emission is derived by fitting to a model discussed in the main text. The black lines are guides for the eyes.

The charge trapping process usually involves many-body interactions between electrons and phonons. It has been found that electron-phonon coupling makes significant contributions to the emission broadening [46] and charge transport scattering [47] in perovskites. Importantly, it has also been suggested that the orthorhombic-to-tetragonal phase transition would not induce significant changes to the phonon spectrum,[48] the carrier-phonon scattering rate,[47] or the phonon energy.[46] Thus, the phase-dependent exciton trapping behavior is unlikely to be caused by phonon variations and more likely to be related to the underlying electronic properties. An indirect bandgap scenario has been put forward to explain the slow and temperature-dependent second-order recombination of perovskites.[26] However, it is difficult for this model to explain the observed ultra-low trap capture cross-sections, particularly since the trap states are usually non-localized in the $k$-space, and there is little evidence of an indirect bandgap. In addition, no obvious differences in the Rashba splitting were found for different perovskite phases,[49] suggesting that a direct-indirect bandgap transition also did not accompany the phase transition.

Inspired by the multiphonon transition theory for solids and semiconductors,[44, 50] here we propose that the phase transition may influence the electronic potential of the point trap in the lattice-coordinate ($Q$) space by changing the coupling between the continuum band and trap states (see Figure 5(c) and Supplementary material note 2). In the non-radiative charge capture process, the total energy ($U$) is distributed between electronic and lattice contributions ($U$=electronic + elastic energy). The charge has to transfer at the state coupling point $Q_O$ with an energy barrier of $E_{B(O)}$ - the energy difference between the bottom of the conduction band and the state coupling point. The phase transition would change the lattice symmetry, atom spacing, and rotational freedoms.[51-52] These changes in atomic and lattice structure have been shown to decrease the elastic constants of perovskites after phase transition,[53] indicating a weakening of the electron-lattice interaction. This would cause a trap state in $Q$ space to shift from $U_{t(O)}$ to $U_{t(T)}$, while also shifting the state coupling point to a higher energy position ($Q_T$) to the bottom of the conduction band. The charge capture energy barrier is thus increased to $E_{B(T)}$, which would significantly reduce the charge capturing probability.[44] This multiphonon transition model provides a qualitative and self-consistent explanation for the phase-dependent exciton trapping behavior, but further experimental support is needed to verify its applicability.

Within this model, following exciton trapping, the captured charges could recombine vertically according to the Franck-Condon principle to emit a photon, i.e. trap emission.[44] This has been observed by us across the whole temperature region below the phase transition. According to the configuration coordinate diagram depicted in Figure 5(c), the increase in the energy barrier $E_B$ is expected to be accompanied by the movement of the trap emission path. The vertical energy difference ($\Delta E_0$) between the bottom of the trap state and the valence band is thus supposed to be increased, which could be reflected by the PL peak energy ($\Delta E$) in experiment. However, the significant dependence between the PL peak energy and the excitation intensity (see Figure 5(d) and Supplementary Figure 10) makes it difficult to directly extract the temperature-dependent $\Delta E_0$. Through extrapolation, we find a logarithmic correlation between the $\Delta E$ and the integrated $I_{PL}$ of trap emission under low excitation densities of $10^{13}$ to $10^{16}$ cm$^{-3}$ (Figure 5(e)). The modeling of trap occupation (Supplementary material note 3) provides a good fit to the experimental results and permits $\Delta E_0$ to be readily extracted (see Figure 5(e)). For temperatures lower than 110 K, $\Delta E_0$ is in the range of 1.64 eV and 1.66 eV, which is also directly reflected by the overlapping of the $\Delta E$-$I_{PL}$ plots. For 120 K and 130 K, the $\Delta E_0$ is significantly increased to 1.72 eV and 1.77 eV, respectively. This giant increase in $\Delta E_0$ of more than 100 meV is strong evidence to support an increase in $E_B$ and, consequently, for the multiphonon transition scenario proposed. This suggests that the charge trapping reduction with the tetragonal phase originates from the increase in the charge capture activation energy following the phase transition.

## Conclusions

In conclusion, the sub-gap energy state related absorption and emission characteristics of lead halide perovskites have been systematically studied in this work to derive the exciton- and trap-state and their dynamic properties. The coexistence of free and bound exciton states have been discovered in both the perovskite polycrystalline and single crystal (both surface and bulk), which may help broaden the perovskite application as exciton devices. The exciton emission and trapping processes are closely related to the trap states and exhibit significant phase-dependent behavior. In the orthorhombic phase, the excitons can be trapped within picoseconds, resulting in strong trap emission. In the tetragonal phase, this process is prohibited, yielding an ultralow trap capture cross section of $10^{-18}$ $cm^2$ and long carrier lifetime up to several hundred nanoseconds. This behavior has been explained within the multiphonon transition theory, stemming from the increase of trap capture activation energy, likely due to a reduction of electron-lattice interaction following the orthorhombic-to-tetragonal phase transition. These findings provide a deeper insight into the distinctive photophysical properties of the perovskite class of material, as well as provide microstructural correlations to enable the development of optoelectronic devices with tailored spectroscopic signatures and, ultimately, higher performance.

## Methods

**Perovskite sample preparation.** For the polycrystalline perovskite powder dispersion, it was formed by stabilizing the perovskite powders in the viscous ethylacetate (EA)/ethylcellulose (EC) solution. Firstly, 0.2 g ethylcellulose (Sigma-Aldrich 46060 and 46070) was dissolved in 15 ml ethylacetate by stirring at 65 °C for 1 hour in a closed bottle. For the $MAPbBr_3$ ($MAPbI_3$), 0.2 M $PbBr_2$ ($PbI_2$) and 0.2 M MABr (MAI) were sequentially poured into the viscous EA/EC solution under the vigorous stirring and heating. The solution became bright yellow as soon as the perovskite raw materials contacted with each other. With continuous heating and stirring, the color became darker. After 2 hours, the raw materials were completely transformed into the perovskite polycrystalline, which was evenly dispersed and stabilized in the solution. For the dilution process, 50 ml EA/EC solution was prepared at first. For a certain perovskite concentration, 1 ml 0.2 M perovskite was diluted by the pure EA/EC solution under the stirring condition.

The $MAPbBr_3$ polycrystalline film (~200 nm) was prepared by the anti-solvent method, where the quartz glass was used as the substrate. 1 M $MAPbBr_3$ dissolved in the DMF was used as the precursor. For film deposition, the $MAPbBr_3$ solution was dropped onto the substrate and spun coated at 5000 rpm, and 100 ml of chlorobenzene was dropped onto the film surface rapidly 5 s after the beginning of spin coating. The final smooth film was heat treated at 100 °C for 10 min. After that, 2 mg/ml PMMA solution (chlorobenzene) was coated onto the film by the spin coating process.

The $MAPbBr_3$ single crystal was grown by the inverse temperature crystallization method as previously reported. Firstly, 1 M $MAPbBr_3$ solution (DMF) was prepared by the continuous stirring at room temperature for 2 h. The precursor solution was poured into a clean bottle with a diameter of about 3 cm through two series-connected 0.45 μm filters. This closed bottle was then put into the oil bath and the temperature was gradually increased and stabilized at 80 °C. After about 2 h without any disturbance, two to three small cuboid crystal nucleuses appeared. With another 2 h, the crystals grew freely and a size of several millimeters was obtained. The single crystal with a well shape was chosen as the sample, which was taken out from the bottle and put into the 2 mg/ml PMMA solution (chlorobenzene) rapidly. This single crystal was stored in the PMMA solution until used. Just before the optical measurements, the single crystal was also cleaved in the glovebox by a sharp scalpel under a certain pressure. The cutting plane is parallel to the top surface of the crystal, which is (100) plane of the cubic $MAPbBr_3$. An optical flat surface without any physical cracks can be obtained by this cleaving process. Immediately after the cleaving, the targeted crystal was re-put into the PMMA solution and then taken out for the following measurements.

**Optical measurements.** For the possible structure relaxation, the perovskite samples, except the cleaved bulk single crystal, were kept in the dark for one day before all the optical measurements. The quartz cuvette is used to store the perovskite dispersion for light absorption measurements, which are conducted by a UV-vis spectrometer (Shimadzu 3600). The steady-state and transient PL of the perovskite dispersion, film and single crystal were all measured by the Edinburgh Fluorescence Spectrometer (FLS 920). A 445 nm pulsed diode laser (EPL-445, ~5 nJ $cm^{-2}$, 62 ps) was used as the excitation source. A circle adjustable neutral density filter was adopted to adjust the excitation intensity. The PL was collected by the reflectance mode, where a PMT together with a TCSPC module was applied to detect the time-resolved or time-integrated PL. For the temperature dependent measurements, an ARS liquid helium cryostat was employed as the sample chamber, where the temperature was controlled by the Lakeshore controller. The perovskite film on the glass substrate was directly fixed to the optical sample holder by screws while the single crystal was pasted to a glass substrate by the high-thermal-conductive low-temperature vacuum grease (Apiezon N). During the temperature adjustment, the system was firstly stabilized for 5 min at each temperature. Before the PL measurements, the perovskite sample was always kept in dark without any bias light or laser illumination to avoid the possible production of charge accumulation and local electric field. For the trap density measurements, a high-energy OPO laser (Opotek Vibrant, 410 nm) was used as the excitation source. A series of fixed neutral density filters and a circle adjustable neutral density filter (Zolix) were applied to adjust the pump fluence at the sample, which was calibrated by a laser energy meter (OPHIR, NOVA II). The PL was measured by a reflectance mode, which was focused into an optical fiber and collected by a CMOS spectrometer (Avantes, EVO) in a time-integrated mode.

**Averaged photocharge density (*n*) calculations.** For the trap density measurement at room temperature, the perovskite PL was measured in the pulse excitation mode, where the laser repetition time is much longer than the charge lifetime of the perovskite. Thus, the photocharge density can be estimated as *n*=excitation fluence each pulse/(photon energy × optical penetration depth). The PL photons are estimated to have a penetration depth of 200 nm for both the thin film and single crystal. For the observation of trap emission blue-shift at low temperature, the carrier lifetime is much longer than the laser repetition time. The perovskite film was excited in an approximately steady-state mode. Thus, a relationship exists as $G_{exc}=n/\tau_0$, where $G_{exc}$ is the generation rate of non-equilibrium charge derived from the excitation power and $\tau_0$ is the charge recombination lifetime.

## Acknowledgements

This work was supported by the Natural Science Foundation of China (Grant Nos. 51372270, 11474333, 51421002, 91433205, and 51627803). JJ would like to acknowledge funding through the ARC Centre of Excellence in Exciton Science (CE170100026).